\documentclass[a4paper,fleqn,usenatbib]{mnras}

\usepackage{graphicx}
\usepackage{times}

\title[Formation of wide binary stars]{Formation of wide binary stars from adjacent cores }

\author[Tokovinin]{Andrei Tokovinin\thanks{E-mail: atokovinin@ctio.noao.edu} \\
Cerro Tololo Inter-American Observatory, Casilla 603, La Serena, Chile}

\date{Accepted XXX. Received YYY; in original form ZZZ}
\pubyear{2017}

\begin{document}
\label{firstpage}
\pagerange{\pageref{firstpage}--\pageref{lastpage}} 

\maketitle

\begin{abstract}
Wide gravitationally bound pairs of  stars can be formed from adjacent
prestellar cores  that happen to  move slowly enough relative  to each
other. These  binaries are remnants of the  primordial clustering.  It
is shown that the expected fraction of wide bound pairs in low-density
star  formation  regions can  be  larger  than  the fraction  of  wide
pairs in the field. On the other hand, wide binaries do not form or
survive   in  dense   clusters.   Recent   works  on   the  separation
distribution  of  young   binaries,  summarized  here,  confirm  these
expectations. Alternative  formation mechanisms of  wide binaries such
as cluster dissolution or unfolding of triple stars cannot explain the
large observed fraction  of young wide pairs and  therefore are not
dominant.   The fact  that  more than  a  half of  wide pairs  contain
subsystems matches  the general  multiplicity statistics and  does not
imply that hierarchical multiplicity and wide binaries are genetically
related.
\end{abstract} 


\section{Introduction}
\label{sec:intro}

Stars form  by fragmentation and collapse of  molecular clouds.  Dense
cores of  molecular gas are believed  to be the  intermediate stage in
this process, with each core forming one or several stars and the core
mass  function being  similar  to the  stellar  initial mass  function
\citep{Alves2007}.  Multiple stars form  by fragmentation of cores and
help to  get rid of  excessive angular momentum  by storing it  in the
orbital motion \citep{PP6}.

\begin{figure}
\includegraphics[width=\columnwidth]{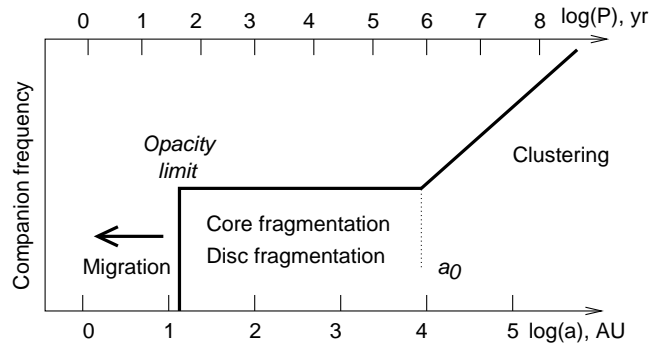}
\caption{Characteristic  scales of binary  formation.  The  thick line
  shows schematically the number  of stellar pairs vs. separation (on
  the lower axis) or period (on the upper axis). Fragmentation creates
  pairs  with separations  from 10  to $10^4$  AU, and  some  of those
  migrate to  smaller separations. At  the spatial scales  larger than
  $10^4$ AU, the  original fractal clustering of stars  persists for a
  long time. The majority of  very wide pairs in the clustering regime
  are not bound, but some are.
\label{fig:cartoon} }
\end{figure}

The typical  separation of  binaries formed by  a fragmenting  core is
related to  its size (on the order  of $10^4$ AU) and  rotation and is
somewhere  between 10  and 1000  AU \citep{Sterzik2003},  matching the
simulations by  \citet{Lomax2015}.  The smallest  separation  on
the  order  of  10~AU  is  set  by  the  so-called  opacity  limit  to
fragmentation  \citep[e.g.][Sect.   4.1]{Goodwin2007}.   Newly  formed
binaries  continue  to accrete  gas  and  at  the same  time  migrate,
evolving      into     tighter      pairs     or      even     merging
(Figure~\ref{fig:cartoon}).

Stellar pairs with  the semimajor axis $a$ of $10^4$  AU or wider
are known  to exist in  the field \citep{Makarov2008}. As  the typical
size of the  cores is similar, on the order of  the Jeans length, such
wide  binaries cannot  be formed  by fragmentation  of a  single core.
Special wide-binary  formation mechanisms have been  proposed, such as
cluster     dissolution     \citep{Kouwenhoven2010,Moeckel2010}     or
``unfolding''  of  more compact  dynamically  unstable triple  systems
\citep{RM12}.

I  explore  the  hypothesis  that  the  components  of  wide  binaries
originate from  different cores.  Owing  to the natural  clustering of
the newly born stars, wide pairs of such stars are very frequent. Some
of  them happen  to  be bound  accidentally.  The
estimated fraction of the wide bound pairs is larger than the fraction
of  wide  pairs  in  the  field  and  even  in  the  young  stellar
groups. Recently  \citet{Joncour2017} discovered pairs  of young stars
in  Taurus with separations  up to  60\,kAU and  argued that  they are
remnants of the original clustering.

The idea  of wide binaries forming from  two independent condensations
has  a long  history. It  was invoked  by \citet{AL76}  to  explain the
different  mass  ratio  distributions  in wide  and  close  solar-type
binaries.   These  distributions  were  revised by  \citet{DM91},  who
nevertheless also found statistical differences between wide and close
pairs  and discussed  ``independent condensation''  of  wide binaries.
More  recent studies  have shown  that  the mass  ratio of  solar-type
binaries depends on the period weakly, if at all \citep{R10,Tok2014}.

\citet{DK13}  discuss  ``extremely wide  systems''  in Section~5.5  of
their review, defining them as binaries with separation 10 times wider
than the  (uncertain) empirical mass-dependent  upper separation limit
(about  $10^4$ AU for  one solar  mass).  Another  way to  define wide
binaries is by  comparing the accretion time with  the orbital period.
If a  protostar of one  solar mass is  assembled in $10^5$  years, the
orbital period of a binary with $a > 3$ kAU exceeds the star formation time.

Wide  binaries  move so  slowly  that  it  is extremely  difficult  or
impossible  to test observationally  whether they  are gravitationally
bound or  just have common spatial motion  and distance \citep[Proxima
  Cen  is a  good example,][]{Kervella2017}.   Frequent  occurrence of
subsystems  further complicates  such  tests. The  term ``wide  pair''
refers here  to a  pair of related,  but not necessarily  bound stars,
reflecting  this  observational   uncertainty.   In  contrast,  ``wide
binary'' means a bound system. 

In  Section~2, we  show that  the clustering  and kinematics  of young
stars allows formation of a substantial number of wide bound binaries.
Many of them contain closer subsystems because in a wide orbit there is
a sufficient range of  allowable stable periods around each component.
The observed  separation distribution and fractions of  wide pairs are
reviewed  in  Section   3;  there  are  many  such   pairs  in  sparse
environments,  while  in dense  clusters  there  are  few. Section  4
discusses  alternative mechanisms  of forming  wide binaries  that has
been proposed in the literature. Section 5 is the summary.

\section{Wide binaries and clustering}
\label{sec:obs}

\subsection{The broken power law}

\citet{Larson1995} discussed  the distribution of  separations between
stars in the Taurus-Auriga  star-forming region (SFR) by combining the
study  of the  large-scale  clustering by  \citet{Gomez1993} with  the
statistics of close binary companions.   He found a break in the power
law  that   approximates  the  density  of   companions  vs.   angular
separation  $\theta$.  At projected  separations larger  than 8250~AU,
the  density of  companions per  square degree  follows the  power law
$\Sigma_c = 3.4 \theta^{-0.62}$.   This corresponds to the logarithmic
separation distribution $f(\log  s) \propto s^{1.38}$.  Larson relates
this to the fractal structure of the stellar clustering with a fractal
dimension  $D \approx 1.4$,  inherited from  the fractal  structure of
molecular clouds.   On the other  hand, at separations between  10 and
8250  AU  the  surface  density  of  companions  in  Taurus-Auriga  is
$\Sigma_c = 0.0064 \theta^{-2.15}$,  or $f(\log s) \propto s^{-0.15}$.
The distribution of  $\log s$ is almost flat  and matches the \"Opik's
law for  binaries.  This range  of (small) separations  corresponds to
binary stars.   The total number  of {\it binary} companions  is about
one per star. 

\citet{Simon1997} repeated  the Larson's study and extended  it to the
Ophiuchus  and Orion  Nebular Cluster  (ONC)  SFRs.  In  all SFRs  the
companion statistics is well represented  by the broken power law. The
break point in  the ONC is found at smaller  separation around 400 AU.
Simon  argues that the  transition between  the clustering  and binary
regimes  occurs  at  separation  on  the order  of  average  projected
distance    between    the    stars,    i.e.    at    the    confusion
limit. \citet{Bate1998} confirm this  by simulations and note that the
power  law at  large scales  can have  multiple origins  and  does not
necessarily  mean   fractal  clustering.    For  a  sparse   SFR  like
Taurus-Auriga,  they find  that  the  break in  the  power law  indeed
corresponds to the average  spatial separation between the stars which
is on the order of the Jeans length.

However, the  transition between the clustering and  binary regimes is
smooth,  stars formed  in  adjacent  cores can  ``fall''  to a  common
centre,  interact dynamically,  and produce  a bound  binary  if their
relative velocity is  small enough. We estimate below  the fraction of
potentially bound wide pairs.  \citet{Kraus2008} suggested that the binary
and clustering regimes are separated by a small region with a constant
density  of  companions  where  the primordial  clustering  has  been
destroyed by motions of the stars. \citet{Joncour2017} explored 
statistics of wide pairs in Taurus by several alternative methods and
found a population of wide pairs with an approximately log-flat separation
distribution extending to 60\,kAU, with preferentially coeval components.

\subsection{Kinematics of cores and binaries}
\label{sec:vel}

\begin{figure}
\includegraphics[width=\columnwidth]{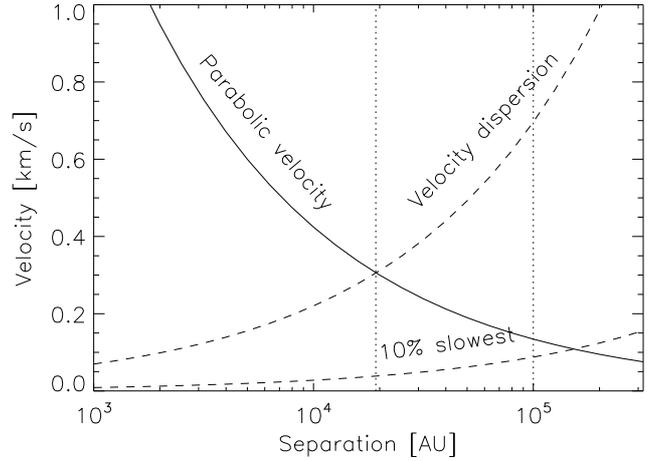}
\caption{Dependence of the parabolic  velocity for a solar-mass binary
  on  its separation  and  the velocity  dispersion  in the  molecular
  clouds.   The lower  curve shows  the velocity  of the  10\% slowest
  stars,  assuming normal  distribution.  At  separation of  $10^5$ AU
  (0.5  pc), the  dispersion  is  5 times  larger  than the  parabolic
  velocity, but  about 15\% of  star pairs still  can be bound  if the
  relative velocity of the cores is normally distributed.
\label{fig:veldiag} }
\end{figure}

Stars form in groups. The  spatial locations of the newly formed stars
and their  velocities are inherited  from the parent  molecular cloud.
Recent studies  have shown  that the gas  is organized  in filamentary
structures.   The diameters  of  the  filaments are  on  the order  of
0.1\,pc, with a remarkably  small dispersion around this typical value
\citep{Andre2014}.   Prestellar  cores  form  along the  filaments  in
chain-like  linear  configurations, with  typical  separations on  the
order of  the filament diameter.  Linear  configurations correspond to
the  fractal  dimension $D  \sim  1$. The  number  of  stars within  a
distance $R$  grows approximately as  $R^D$.  \citet{Larson1995} found
$D  \approx 1.4$  for the  Taurus-Auriga SFR,  later \citet{Kraus2008}
revised it to $D \approx 1$.

Gas motions  in molecular  clouds are described  by the  Larson's law,
where  the  typical  rms  velocity difference  between  two  fragments
$\sigma_V$  is  roughly  proportional  to  the square  root  of  their
separation    $s$,   $\sigma_V(s)    \approx    (s/1{\rm   pc})^{1/2}$
[km~s$^{-1}$]  \citep{Heyer2009}.   On  the  other hand,  the  orbital
velocity of a binary decreases  with its semimajor axis $a$ as $V_{\rm
  orb} = 30 a^{-1/2} M^{1/2}$ [km~s$^{-1}$], $M$ being the mass sum in
solar units and $a$ ---  semimajor axis in AU.  This formula describes
the  differential velocity in  a circular  orbit, while  the parabolic
velocity $V_{\rm par} = \sqrt{2}  V_{\rm orb}$. A pair of stars moving
faster  than  $V_{\rm par}$  has  positive  total  energy and  is  not
gravitationally bound.

Suppose that  two gas  fragments of a  collapsing cloud form  a binary
with $a  \approx s$, thus storing  the angular momentum  in its orbit.
Then the condition $\sigma_V \le V_{\rm par}$ must hold, otherwise the
fragments   will  not   be   gravitationally  bound   to  each   other
(Figure~\ref{fig:veldiag}).  This leads to
\begin{equation}
s < a_0 =  1.9 \, 10^4  M ^{1/2} {\rm [AU]},   
\label{eq:amax}
\end{equation}
or  $a_0 \approx 0.1$\,pc. The maximum separation  of a  solar-mass binary
$a_0$ is  therefore defined by the gas  motions to be on  the order of
$10^4$ AU.   This is  also the characteristic  size of  the prestellar
cores  and of  the same  order  as the  Jeans length  and the  typical
diameter  of filaments  in  molecular clouds.   Note  that this  scale
depends on   mass; it is smaller for low-mass  cores and larger for
massive cores.  Environments with faster gas motion should produce, on
average, closer binaries.

\subsection{Fraction of wide bound  pairs}
\label{sec:frac}

Cores separated by the distance   $s > a_0$ on average move too fast
to form a gravitationally bound pair. However, the motions are chaotic
and a certain fraction of adjacent cores still happen to be bound. 

Suppose that  the relative velocity  of cores is  distributed normally
with a dispersion $\sigma_V(s) \propto s^{1/2}$. The fraction of cores
with  a  relative  velocity   less  than  $V$  is  then  approximately
$\sqrt{2/\pi} V/\sigma_V = 0.8 V/\sigma_V$, assuming $V \ll \sigma_V$.
For bound  pairs, we  require $V  < V_{\rm par}$,  so the  fraction of
bound core  pairs among all neighbors should  decrease with separation
as $0.8 V_{\rm par}/\sigma_V \propto s^{-1}$.

The  initial clustering  of young  stars  is such  that the  companion
density is  $f(s) \propto  s^{0.4}$ (assuming $D=1.4$).   The expected
fraction  of  bound  pairs  with  $s \gg  a_0$  is  therefore  $f_{\rm
  bound}(s)   \propto  s^{-0.6}$.    The   log-flat  distribution   of
separations  corresponds  to  $f(s)  \propto  s^{-1}$,  so  the  above
argument  predicts the number  of wide  bound pairs  in excess  of the
\"Opik's law, or matching it for $D \approx 1$.

If the relative motions of  the cores are approximately isotropic, the
modulus of the  relative velocity $V$ is distributed  according to the
Maxwell-Bolzmann law
\begin{equation}
f_{\rm Maxwell}(V)  = \sqrt{2/\pi} \;  V^2 \sigma^{-3} \exp (  - V^2/2
\sigma^2 ).
\label{eq:Maxwell}
\end{equation}
In such case, the fraction of slow movers (bound pairs) decreases with
$s$ much  faster, as  $(V/\sigma_V)^3 \propto s^{-3}$.   Therefore, in
the case  of isotropic  relative motions the  number of  wide binaries
than can be formed from adjacent cores drops sharply at $ s > a_0$.

It is conceivable that in  the filamentary clouds the relative motions
are   highly  anisotropic,   with  one   prevailing   direction  (e.g.
quasi-rotation).  If  motions in one direction  dominate, the Gaussian
distribution of the relative velocity is a fair approximation, and the
number of wide bound pairs can be significant. Bound wide binaries can
also  originate  from regions  with a slower  than  average relative  gas
motion. Decreasing the velocity dispersion by a factor of two doubles the
$a_0$, extending it to 40\,kAU.

\subsection{Time scale}
\label{sec:time}

The orbital period of a binary is related to its semimajor axis by the
third Kepler law, $ P =  a^{2/3} M^{-1/2}$, where $P$ is in years, $a$
is in AU  and $M$ is the mass  sum in solar units. The  upper scale in
Figure~\ref{fig:cartoon}  shows  the  period  vs.   separation  for  a
solar-mass  binary.    The  characteristic  scale  $a_0   =  10^4$  AU
corresponds to $P=1$\,Myr.  The time  needed for the two cores to fall
to the  centre of mass or to  ``unfold'' is about $P/2$.   If the star
formation and  accretion last for  about $10^6$ years,  the pair  of cores
separated further  than $a_0$ has not  yet had time to  fall onto each
other  and  cannot  be considered  as  a  binary,  even if  they  will
eventually become bound. 

The Larson's law predicts  faster motions at larger separations; e.g.,
two stars  separated by 1\,pc  move at $\sim$1\,km~s$^{-1}$  speed and
will travel 1\,pc in 1\,Myr.   This primordial motion is not generally
directed toward the centre of mass and most likely will move the stars
apart.  The $\sigma_V(s) \propto s^{1/2}$ dependence means that larger
initial configurations will be preserved longer; the clustering at the
spatial  scale  $s$ will  persist  for  $t  \sim (s/1{\rm  pc})^{1/2}$
Myrs.  In other  words,  the  smallest spatial  scale  of the  remnant
clustering is  proportional to $t^2$.   Young objects of Class  0 have
ages on the order of $10^5$ years and keep their primordial clustering
at the scale of 0.01\,pc or 2 kAU, while at 10\,Myr this scale becomes
100\,pc and all clustering  is lost. \citet{Kraus2008} demonstrate how
the original clustering has been destroyed on small scales by modeling
the observed  companion density with  a doubly broken power  law.  The
knee positions in these fits  match the velocity dispersion times age.
A transition  zone between the two  power-law segments was also  found for
young binaries in the Orion molecular clouds by \citet{Kounkel2016}.

By similar argument, tight pairs should form  and evolve faster than
the wide  ones.  The free-fall time is  proportional to $\rho^{-1/2}$,
so   dense   regions   collapse   in  a   runaway   manner,   creating
singularities. However, those star embryos  have only a small mass and
acquire most  of their  final mass by  accreting the  surrounding gas.
\citet{Larson2007}   notes  that  ``self-gravitating   structures  may
generally  be   built  or  organized   from  the  bottom   up  because
gravitational processes operate faster  in smaller and denser regions,
so that  the matter  is collected together  first on small  scales and
then on progressively larger  scales''. This means that multiple stars
form mostly  ``from inside out'',  starting with the  inner subsystems
and adding  outer components.  When the  two cores combine  in a bound
wide binary, some of them  already contain subsystems.  We expect 
the orbits of those subsystems to be randomly aligned  relative to each
other and to the outer orbit.

\subsection{Why wide pairs often contain subsystems}
\label{sec:mult}

It  has been  noted that  many  wide pairs  contain inner  subsystems,
i.e. are hierarchical triple, quadruple or higher-order multiples. For
example, \citet{Elliott2016} found that in the $\beta$~Pictoris moving
group (BPMG), 11 out of  of 14 pairs with projected separations larger
than $10^3$ AU are hierarchical multiples.  \citet{Law2010} found that
45\%  of  wide M-dwarf  pairs  are  hierarchical  multiples, with  the
fraction   of   hierarchies  being   larger   at  wider   separations.
\citet{Joncour2017} estimate  that $\sim$68\% of wide  pairs in Taurus
with $s > 1$\,kAU  contain subsystems.  High incidence of hierarchical
systems  with  wide outer  separations  has  been  put forward  as  an
argument for  preferential formation of wide binaries  by unfolding of
more compact triples \citep{RM12}.

The statistics of solar-mass  hierarchical multiple stars in the field
can be modeled  as an almost independent combination  of the inner and
outer  subsystems  drawn  from  the same  generating  distribution  of
periods   and    restricted   only   by    the   dynamical   stability
\citep{Tok2014}. In a wide pair, the range of dynamically stable inner
separations is large,  hence the incidence of subsystems increases
with the outer separation.  Suppose that this range corresponds to the
subsystem probability $\epsilon =  0.5$. The fraction of wide pairs
without subsystems  is then $(1  - \epsilon)^2 = 0.25$,  the remaining
75\% of  wide pairs are  hierarchical multiples.  This  corresponds to
10.5 hierarchies among the 14 wide pairs in the BPMG.

The  field  multiplicity model  \citep{Tok2014}  does  predict a  mild
correlation between  the wide (outer) and close  (inner) subsystems by
postulating that the field is a mixture of binary-rich and binary-poor
populations.  Wide  binaries come from the low-density  SFRs that also
have  an  increased  binary  fraction,  hence  a  larger  fraction  of
subsystems.   Moreover, both  \citet{Law2010}  and \citet{Joncour2017}
note  that wide  pairs containing  subsystems have  larger  masses and
larger binding  energies compared to pure wide  binaries, helping them
to survive.   The environment  effect and the  binding-energy argument
may be sufficient for  explaining the fraction and period distribution
of  hierarchies  in  the  field  without  assuming  special  formation
mechanisms of wide binaries related to hierarchical multiplicity.

\section{Observed frequency of wide pairs}

\subsection{Low-density SFRs}

\begin{table}
\centering
\caption{Distribution of projected separations 
\label{tab:sephist} }
\begin{tabular}{l cccc}
\hline
$\log s$ (AU) & Perseus$^a$ & Taurus$^b$ & BPMG$^c$ & Simulations $^d$ \\
\hline
($-$0.5, 0) & \ldots  & \ldots & 5   & 7  \\
(0, 0.5)    & \ldots  & \ldots & 4   & 23 \\
(0.5, 1)    & \ldots  & 18     & 5   & 35 \\ 
(1, 1.5)    & 3       & 17     & 6   & 44 \\
(1.5, 2)    & 8       & 24     & 5   & 22 \\
(2, 2.5)    & 5       & 12     & 3   & 24 \\
(2.5, 3)    & 5       & 18     & 4   & 11 \\ 
(3, 3.5)    & 9       & 10     & 7   & 2 \\
(3.5, 4)    & 13      & 13     & 0   & 5 \\
(4, 4.5)    & \ldots  & 23     & 4   & 2 \\
(4.5, 5)    & \ldots  & 93     & 3   & 0 \\
$N_{\rm tot}$ & 52     & 118    & 49  & 426 \\
\hline
\end{tabular}  \\
\raggedright
References: $^a$ \citet{Tobin2016};  $^b$ \citet{Joncour2017}; $^c$
\citet{Elliott2016}; $^d$ \citet{Bate2014}.  
\end{table}

\begin{figure}
\includegraphics[width=\columnwidth]{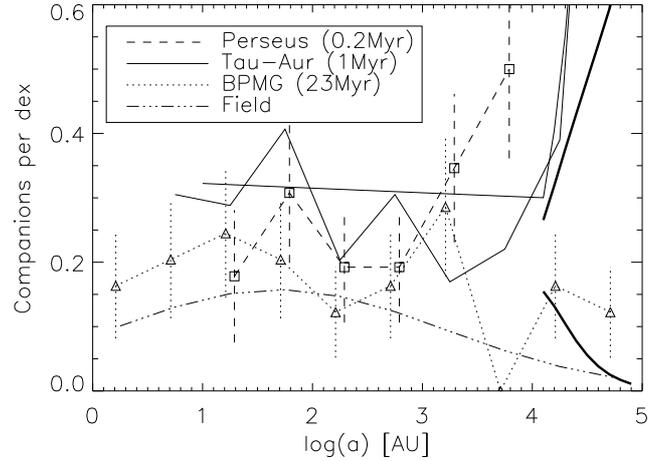}
\caption{Distributions  of projected  separations in  the  Perseus SFR
  \citep[dashed line,][]{Tobin2016},     Taurus-Auriga
  \citep[two-segment and broken full lines,][]{Simon1997,Joncour2017},     BPMG
  \citep[dotted line,][]{Elliott2016},  and  the  field
  \citep[dash-dot line,][]{R10}.    Companion
  frequency per decade of separation  is plotted on the vertical axis.
  The thick solid  lines at $\log a > 4$ are the clustering distributions  scaled down
  according to the expected proportion of wide bound pairs under the
  assumptions of Gaussian (upper) or Maxwell (lower) velocity distributions.
\label{fig:sephist} }
\end{figure}

Binarity of very young protostars in the Perseus SFR has been recently
explored  by \citet{Tobin2016}  using high-resolution  observations at
centimeter  wavelengths.  Their  resolution limit  corresponds  to the
projected separation $s \sim 15$~AU. The distribution of $s$ for Class
0 and Class I sources from their  Tables 3 and 4 is reproduced here in
Table~\ref{tab:sephist} as the number of pairs per 0.5 dex separation bin.
The  last line  of this  Table  gives the  total size  of the  sample,
allowing us to compute the companion frequency. The age of all sources
is less than 0.5\,Myr, the majority of the Class~0 sources are younger
than  $\sim$160\,Myr. 

Table~\ref{tab:sephist}  also gives  the latest  binary  statistics in
Taurus  based on  the Table~C1 from  \citet{Joncour2017}. I  selected from
their catalogue 142 stars with  mass above 0.3 ${\cal M}_\odot$ observed
with high angular resolution, and  computed the number of pairs in each
separation bin,  counting each  pair only once.   To match  the common
convention, 24 pairs of catalog entries within 10 kAU of each other
are  considered as  binary systems,  reducing the  sample size  to 118
(this adjustment  affects only the normalization of  the curve).  The
full  broken line without  error bars  in Figure~\ref{fig:sephist}  is a
good   match   to  the   broken   power   law  of   \citet{Simon1997}.
\citet{Kraus2011} also studied  the separation distribution in Taurus,
but only out to 5\,kAU, avoiding the clustering regime.

Young moving groups  have ages between 10 and 100 Myr;  the age of the
$\beta$~Pictoris moving  group (BPMG) is  23\,Myr \citep{Mamajek2014}.
\citet{Elliott2016} studied the binary  statistics in this group using
a  variety  of techniques.   The  sample  size  is 49,  the  companion
frequency is about one. Data from their Table~2 are used for computing
the histogram  given in  the fourth column  of Table~\ref{tab:sephist}.
\citet{Kraus2009}  derived a  similar logarithmically  flat separation
distributions in  the range  $5 < s  < 5000$  AU for binaries  in both
Taurus-Auriga and Upper Scorpius SFRs.

Binary statistics of the older  field stars of approximately one solar
mass is well studied  \citep{R10,Tok2014}. The distribution of periods
and separations is approximately log-normal, with the median period of
$10^5$  days,  logarithmic period  dispersion  of  2.3, and  companion
fraction  about  0.60.  At  $s>3$  kAU,  the  frequency of  companions
declines  approximately  as  a   power  law  $f(s)  \propto  s^{-1.5}$
\citep{TL12}; the  power law fits the  data as well  as the log-normal
model.   About  2\% of  solar-type  stars  in  the field  have  
companions  with $s  > 10^4$  AU.  Figure~\ref{fig:sephist}  plots the
fraction  of   companions  per   decade  of  separation.    The  field
distribution is shown by the dash-dot line, while the full line is the
broken  power law  for  Taurus-Auriga \citep[data  from][]{Simon1997}.
The  clustering power law  scaled  down according  to the  expected
fraction of  bound pairs  (thick solid lines)  shows the  frequency of
potentially bound wide pairs originating from adjacent cores under two
assumptions (normal or Maxwell distributions of the relative velocity).

In the youngest  Perseus SFR there are many  pairs with separations of
$\sim10^3$ AU and wider, while  at smaller separations the numbers are
approximately   constant,    within   the   errors    (a   log-uniform
distribution).  A similar  upturn  in the  separation distribution  of
protostars  at   $s  >  10^{3.5}$   AU  has  been  found   earlier  by
\citet{Conneley2008}.   In  the older  Taurus-Auriga  SFR, the  upturn
occurs at separations $s>10^4$ AU,  while in the BPMG the distribution
is  approximately flat  out to  $10^5$  AU (note  however the  large
statistical error  bars).

The multiplicity fraction depends on the mass of the primary component
\citep{DK13}.    Data   on   the   young   associations   plotted   in
Figure~\ref{fig:sephist} refer  to a  mixture of masses,  mostly below
one  solar,  while the  masses  in Perseus  are  not  known and  these
protostars are  still accreting.   The comparison of  the multiplicity
fraction with  the field  (masses around one  solar) should  take this
into account.  The histograms  show that the multiplicity fractions in
Taurus, BPMG  and  Perseus are  larger  than in  the  field  even at  moderate
separations   $ s   \sim  10^2$   AU,   as  in   other  sparse   SFRs
\citep{Kraus2009,King2012}.    Such   binaries   are   not   disrupted
dynamically and survive  in the field.  At the  same time, sparse SFRs
produce  most (or all)  wide binaries.   Not  surprisingly, those  wide
binaries often contain subsystems.

\subsection{Dense SFRs}

\begin{figure}
\includegraphics[width=\columnwidth]{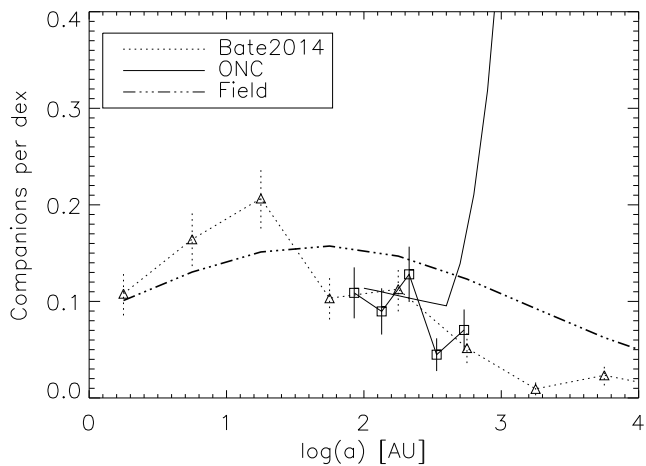}
\caption{Distributions  of projected  separations in  the ONC (full lines), 
in a simulated dense cluster (dotted line), and in the field.
\label{fig:orihist} }
\end{figure}

In  dense regions,  wide binaries  cannot  form or  survive, owing  to
dynamical  interactions with  their neighbors.   The  typical distance
between stars is  less than the core size or the  Jeans length, so the
concept of  core becomes questionable in  these conditions. Protostars
move  in   a  common  gas   cloud  while  accreting   and  interacting
dynamically.   \citet[][and references  therein]{Bate2014}  made large
hydro-dynamical simulations of the collapse of a dense turbulent cloud
containing 500\,$M_\odot$ of gas.  Simulations cover 0.3\,Myr, until a
substantial fraction of the gas is accreted.  Although the filamentary
structure and  fractal-like clustering  of newly formed  stars develop
initially,  the subgroups subsequently  merge, with  violent dynamical
interactions  between  the  stars  and continuing  accretion.   Citing
\citet{Moeckel2010}, ``..the stars form  in a structured fashion, with
smaller sub-clusters merging  to form a final cluster  consisting of a
tightly  bound core  with  radius $\approx$0.05  pc  surrounded by  an
expanding    halo    of    ejected    stars.''     We    provide    in
Table~\ref{tab:sephist}  the distribution of  the semimajor  axes from
\citep{Bate2014}  that  drops  sharply  at $a  >  1$\,kAU.   Subsequent
dynamical evolution of the  cluster may curtail this distribution even
further  \citep{Kaczmarek2011}.   However,  \citet{Moeckel2010}  argue
that after the gas dispersal  the cluster expands and does not destroy
the binaries  that have  already formed in  a denser  environment. The
multiplicity  fraction in  the expanding  cluster  stays approximately
constant, except in its outer halo where it is $\sim$2 times less.

Figure~\ref{fig:orihist}  plots the  separation  distribution for  the
high-density SFR, namely the ONC.  Squares and full line correspond to
the Figure~7  of \citet{Reipurth2007},  who  studied the  binarity in  the
separation range from 67.5 to 675  AU in a sample of 781 stars outside
the  Orion  Trapezium.    The  broken  power  law  is   based  on  the
\citet{Simon1997} fit  to the  companion statistics in  the Trapezium.
It is reduced by a factor 2 to match approximately the binary fraction
found by  Reipurth et  al. Such adjustment  is justified by  the large
errors of  the power-law fits  in \citep{Simon1997} and by  the larger
primary  mass (hence  higher  multiplicity) in  the Trapezium  sample.
\citet{Bate1998}  convincingly  show  that  the  break  point  in  the
power-law fits is  related to the average star  density, which is high
in the  Trapezium.  We do not  see the sharp upturn  in the separation
distribution  of binaries  outside the  Trapezium because  the stellar
density  there   is  much  less,  and  because   the  contribution  of
neighbouring stars has been subtracted from the histogram.

For  comparison, we  plot in  Figure~\ref{fig:orihist}  the separation
distribution  of  binaries resulting  from  the large  hydro-dynamical
simulations  by \citet{Bate2014}  (see  Table~\ref{tab:sephist}).  The
simulated cluster roughly matches the  ONC in density.  Note the sharp
drop of binary frequency at separations beyond 300 AU. There are still
a  few soft  binaries with  $s>10^3$\,AU  that will  likely be  disrupted
within the cluster, while new  wide binaries will appear in its expanding
halo (see Section~\ref{sec:cluster}).

At separations around  100 AU, the companion frequency  in the ONC and
in the simulated cluster is less than for the solar-field dwarfs. 
This statistics refer  to primary stars of less  than one solar mass,
hence such  difference is  expected.  Of importance  here is  the fast
drop in  the number of pairs  wider than $\sim$300  AU, compared to
the separation distribution in the  field and in the low-density SFRs.
\citet{Reipurth2007} further  discuss the paucity of  wide pairs in
the ONC.

\section{Alternative formation mechanisms of wide binaries}
\label{sec:alt}

\subsection{Unfolding}
\label{sec:unfolding}

\citet{RM12} suggested that very wide binaries are formed by ejections
from more  compact unstable triples.  They call  this ``unfolding'' of
triple  systems  into  extreme  hierarchical  architecture.   In  this
mechanism,  the outer orbits  are very  eccentric because  the angular
momentum  of the  wide  pair  is derived  from  the initially  more
compact  system. Although  the  ratio of  semimajor  axes in  unfolded
triples can  be large, they  are still only marginally  stable because
the tertiary and  the binary approach each other  at the periastron of
the eccentric outer orbit.

The unfolding  time is on  the order of  $P/2$. The authors  note that
``Many wide  systems have not  unfolded fully at  1 Myr, and  the most
extreme wide  systems will take tens  to hundreds of  million years to
unfold, and they are thus more protected against disruption by passing
stars.''    In   young  associations   such   as  Taurus-Auriga   (age
$\sim$1\,Myr), there should be no  binaries with $s>10^4$ AU formed by
this  mechanism.   In reality,  young  SFRs  contain  many wide  pairs
(including  unbound  ones),  but  their  fraction  progressively  {\it
  decreases} with age instead of increasing.

In the  simulations by \citet{RM12}, decaying  triple systems generate
about  2\%  of  wide  ($a  >  10^4$ AU)  binaries  per  system,  matching
approximately the  frequency of wide pairs in  the field.  However,
not all  cores fragment  into triple or  quadruple stars, and  not all
triples are  unstable, so  in a realistic  situation the  frequency of
``unfolded'' wide binaries  produced by this mechanism is  too small even
for the field, and much smaller than in young groups such as BPMG.
 
Another  strong  prediction of  the  unfolding  scenario  is the  high
eccentricity of the outer orbits, the  majority with $e> 0.9$ for $a >
10^4$ AU.  There  should be a correlation between  the outer semimajor
axis  and the  outer eccentricity.   The eccentricities  of  very wide
binaries can  be determined  statistically by accurate  measurement of
the relative motion  \citep{TK16}, soon to be available from {\it Gaia}.
However, large incidence of subsystems implies the need to account
for their motion, making such future study quite challenging.

The triple system  $\alpha$~Cen A,B (HIP 71683 and  71681) and Proxima
Cen  (HIP  70890)   is  an  excellent  test  case   for  the  ejection
scenario. Being  the nearest,  it should be  typical of  other similar
systems (rather than exceptional). It is bound, given its age.
The projected separation of Proxima  is $s_{\rm proj} = 10.7$ kAU, the
total separation in space is $s  = 15$ kAU, and the parabolic velocity
is   $V_{\rm  par}   =   0.51$  km~s$^{-1}$.    \citet{Wertheimer2006}
calculated the binding energy of the  wide pair AB,C and found that it
can be either positive or negative with roughly equal probability. The
latest study  by \citet{Kervella2017}, however, shows  that the triple
system  is bound  and  that the  eccentricity  of the  outer orbit  is
moderate, 0.42.  Therefore  this triple system has not  been formed by
unfolding. Interestingly,  the outer and inner orbits  are inclined by
only 29\degr. The outer orbit thus has a large angular momentum and it
is roughly aligned with the inner orbit.

\subsection{Cluster dissolution}
\label{sec:cluster}

\citet{Moeckel2010} show that after  10\,Myr of dynamical evolution of
the dense  simulated cluster, several wide  binaries with $10^4  < s <
10^5$ AU  are formed in its  expanding halo. These  binaries are wider
than the  original cluster  size of  $10^4$ AU, and  half of  them are
triple or higher-order multiples.  There were $15 \pm 7$ binaries with
$s>  10^4$ AU, or  1.7\% for  a total  of 900  stars in  that cluster.
\citet{Kouwenhoven2010}  estimate   that,  on  average,   one  cluster
produces  one wide  binary with  $s>10^4$ AU.   The frequency  of such
pairs in the field is $\sim$2\%.   If all field wide pairs were formed
by the  cluster dispersal mechanism, a typical  cluster should contain
$\sim$50 stars.

On the other hand, the frequency of pairs with $s > 10^4$ AU in the
BPMG is  7/49=0.14$\pm$0.05, significantly  larger than in  the field.
Such a large  fraction of wide pairs cannot  be explained by either
cluster    dissolution    or     by    the    unfolding    mechanisms.
\citet{Joncour2017}  reached  the   same  conclusion  regarding  wide
pairs in Taurus.

\subsection{Entrapment}
\label{sec:entrapment}

\citet{Makarov2012} considered formation of very wide ($s \sim 1$\,pc)
pairs  from unrelated  stars that  are accidentally  ``entraped'' into
common motion in the Galactic  potential. In principle, such pairs can
survive  for a  long  time.   However, their  motion  is chaotic  (not
Keplerian)   and  they   can   dissolve  as   easily   as  they   were
formed. Makarov  estimates that this mechanism is  too inefficient for
explaining wide pairs  in the field. Moreover, wide  binaries are more
readily disrupted  by passing  stars or molecular  clouds than  by the
smooth Galactic potential, so the large Jacoby radius of $\sim$1.8\,pc
is not  relevant for their survival \citep{Weinberg1987}.

\section{Summary}
\label{sec:sum}

The separation  distribution of stellar pairs in  the low-density SFRs
is  nearly uniform  (logarithmically) up  to  $10^4$  AU and
increases at  larger separations.  This upturn is  associated with the
initial  fractal  clustering   of  prestellar  cores.   It  disappears
progressively  with  increasing  age,  leaving  a  smoothly  declining
separation distribution  in the field  (Figure~\ref{fig:sephist}). 
Young  moving   groups  such  as  BPMG  (age   23\,Myr)  represent  an
intermediate  case, with 14\%  of stars  having companions  wider than
$10^4$ AU.  This large fraction of  wide pairs does  not match such
proposed formation  mechanisms as cluster dissolution  or unfolding of
unstable triples.  However, it can be explained easily as a remnant of
the  clustering, where  a  fraction  of adjacent  cores  happen to  be
gravitationally bound.  It is shown that the gas kinematics allows for
a large enough fraction of bound  core pairs if the motions are highly
anisotropic and the resultant velocity difference between the cores is
distributed almost normally, rather than by the Maxwell-Bolzmann law.

Stars in the field come  from different environments. The frequency of
wide pairs in the field therefore can indicate what fraction of the
field stars were formed in the low-density SFRs like Taurus-Auriga and
BPMG,  and  what  fraction  comes  from  larger  and  denser  clusters
\citep{Patience2002,Reipurth2007}.

\section*{Acknowledgments}

I appreciated the discussion  of binary formation with B.~Reipurth and
P.~Elliott. Referee's comments helped to improve this paper.


\label{lastpage}

\end{document}